\documentclass[%
 reprint,
pra,
superscriptaddress,
nofootinbib,
 amsmath,amssymb,
 aps,
]{revtex4-2}

\usepackage{graphicx}
\usepackage{dcolumn}
\usepackage{bm}
\usepackage{braket}  
\usepackage{newtxtext,newtxmath}
\usepackage[hidelinks]{hyperref}

\begin{document}

\preprint{APS/123-QED}

\title{Realignment Criterion: A necessary and sufficient condition for two-qubit $X$-states}

\author{Amish Gogia}
\author{Aakhyat Bagga}
\author{Satyabrata Adhikari}
\author{Mukhtiyar Singh}
\affiliation{Delhi Technological University, Delhi-110042, Delhi, India}

\setlength{\parindent}{0pt}

\begin{abstract}
The Computable Cross Norm (CCN), or realignment criterion, is a widely used method for entanglement detection in quantum systems; however, it typically provides only a necessary condition. In this work, we advance the applicability of the realignment criterion by deriving a condition that is both necessary and sufficient for detecting entanglement in two-qubit. $X$-states derive their name from the characteristic ‘X’ shape of their density matrix, which contains seven independent matrix parameters. Notably, they incorporate several important subclasses of entangled states, including Bell states, Werner states, and maximally entangled mixed states. $X$-states have proven highly useful in entanglement studies due to their sparse structure and the ease with which entanglement-related quantities can be computed. This refined criterion improves the identification of entangled states that the standard CCN approach fails to detect, thereby extending the utility of the method.
\end{abstract}

\maketitle


\section{INTRODUCTION}

Entanglement was first identified by Einstein, Podolsky, and Rosen (EPR), as well as Schrödinger, who famously referred to it as “\textit{spooky action at a distance}” \cite{1}. This phenomenon revealed the existence of quantum composite systems that cannot be described as a product of two independent subsystems. In an attempt to explain this non-classical behaviour, Einstein, Podolsky, and Rosen proposed the existence of hidden variables, suggesting that physical quantities possess well-defined values prior to measurement. However, in 1964, Bell disproved this notion by formulating Bell’s theorem \cite{1a}.

Detecting entanglement remains a fundamental challenge in quantum computing and quantum information theory \cite{2,3}. The ability to distinguish between separable and entangled states is crucial for various quantum technologies. However, entanglement detection is known to be an NP-hard problem \cite{3a}, implying that no single criterion can identify all entangled states efficiently. Over the years, significant research efforts have been devoted to developing various entanglement detection methods \cite{4,5,6,7,8,9,10}.

One of the earliest and most widely used criteria for entanglement detection is the Peres-Horodecki criterion, also known as the positive partial transpose (PPT) criterion \cite{11, 12}. This criterion is both necessary and sufficient for detecting entanglement in $2 \otimes 2$ and $2 \otimes 3$ systems, but in higher-dimensional systems, it is only a necessary condition, meaning that some entangled states may go undetected.

For a bipartite quantum system with Hilbert space \(\mathcal{H}_A \otimes \mathcal{H}_B\), where \(\dim(\mathcal{H}_A) = d_A\) and \(\dim(\mathcal{H}_B) = d_B\), the density matrix $\varrho$ is represented in the standard product basis \(\{|e_i\rangle \otimes |f_j\rangle\}\) as:

\begin{equation}
\label{eq:1}
    \varrho = \sum_{i,j,k,l} \rho_{ij,kl} \, |p_i\rangle \langle p_k| \otimes |q_j\rangle \langle q_l|.
\end{equation}

For a system where \( d_A = d_B = 2 \), the density matrix $\varrho$ is of dimension \( 4 \times 4 \) and can be expressed in block form as:

\begin{equation}
\label{eq:2}
    \varrho_{AB} =
    \begin{pmatrix}
        A & B \\
        B^{\dagger} & C
    \end{pmatrix},
\end{equation}
where each block is a \( 2 \times 2 \) sub-matrix given by:

\begin{equation}
\label{eq:3}
    \begin{aligned}
        A &= \begin{pmatrix} \rho_{11} & \rho_{12} \\ \rho_{12}^* & \rho_{22} \end{pmatrix}, \quad
        B = \begin{pmatrix} \rho_{13} & \rho_{14} \\ \rho_{23} & \rho_{24} \end{pmatrix}, \\
        B^{\dagger} &= \begin{pmatrix} \rho_{13}^* & \rho_{23}^* \\ \rho_{14}^* & \rho_{24}^* \end{pmatrix}, \quad
        C = \begin{pmatrix} \rho_{33} & \rho_{34} \\ \rho_{34}^* & \rho_{44} \end{pmatrix}.
    \end{aligned}
\end{equation}

The partial transposition of $\varrho_{AB}$ may be defined as: 

\begin{equation}
\label{eq:4}
    \varrho_{AB}^{T_B} =
    \begin{pmatrix}
        A^T & B^T \\
        (B^{\dagger})^T & C^T
    \end{pmatrix},
\end{equation}

The realignment operation \cite{13, 14} is another method of systematic rearrangement of these sub-matrices. Each block is vectorized by stacking its elements into a single column vector:

\begin{align}
\label{eq:5and6}
    \text{vec}(A) &= 
    \begin{pmatrix} \rho_{11} \\ \rho_{12} \\ \rho_{12}^* \\ \rho_{22} \end{pmatrix}, \quad
    \text{vec}(B) = 
    \begin{pmatrix} \rho_{13} \\ \rho_{14} \\ \rho_{23} \\ \rho_{24} \end{pmatrix}, \\
    \text{vec}(B^{\dagger}) &= 
    \begin{pmatrix} \rho_{13}^* \\ \rho_{23}^* \\ \rho_{14}^* \\ \rho_{24}^* \end{pmatrix}, \quad
    \text{vec}(C) = 
    \begin{pmatrix} \rho_{33} \\ \rho_{34} \\ \rho_{34}^* \\ \rho_{44} \end{pmatrix}.
\end{align}

The realigned matrix \(\mathcal{R}(\varrho_{AB})\) is then constructed by arranging the transpose of these vectors as the rows of a new matrix:

\begin{equation}
\label{eq:7}
    \mathcal{R}(\varrho_{AB}) =
    \begin{pmatrix}
        \text{vec}(A)^T \\ \text{vec}(B)^T \\ \text{vec}(B^{\dagger})^T \\ \text{vec}(C)^T
    \end{pmatrix}.
\end{equation}

This operation leads us to define a criterion for the detection of entanglement which is stated as follows:

\begin{center}
    \textit{Given a separable state $\varrho_{}$ then the $|| \mathcal{R}(\varrho) ||_1 \leq 1$}
\end{center}

where $||.||_1$ denotes the trace norm and is mathematically defined as $||A||_1$ = $Tr\sqrt{A.A^\dagger}$

In this paper, we focus on redefining the realignment criterion for a particular class of two-qubit states known as $X$-states. The existing realignment criterion serves as a necessary but not sufficient condition for detecting entanglement in two-qubit systems. Our proposed refinement aims to make the realignment criterion as necessary and sufficient criterion and thus making it more effective in identifying entanglement within this specific class of states.

The organization of this paper is as follows: In Section II, we build on the work of \cite{34} and derive a necessary and sufficient condition based on the realignment operation to detect entanglement in two-qubit \( X \)-states. In Section III, we present a generalized two-qubit \( X \)-state that is not fully detected by the standard realignment criterion but is detected over a wider range using the criterion derived in this work. Furthermore, we establish bounds on its density matrix elements for entanglement detection. We finally conclude our findings in Section IV.

\subsection{$X$-states}

Two-qubit $X$-states represent an adaptable category of quantum states, as elucidated in \cite{15}. They include a wide range of some important entangled states, namely, Bell states \cite{16}, Werner states \cite{17}, isotropic states, and maximally entangled mixed states \cite{18, 19, 20, 21}. Their intrinsic properties enable an exhaustive examination of concurrence and purity within bipartite quantum systems, as evidenced by earlier work \cite{22}. Moreover, any arbitrary two-qubit state can be converted into an $X$-state via unitary transformations that conserve concurrence, making them instrumental for quantum state manipulation. Their sparse structure significantly improves both theoretical tractability and computational efficiency, and their experimental realizability has been demonstrated across a range of physical platforms, including nuclear magnetic resonance (NMR), as well as optical and atomic systems \cite{23, 24, 25, 26, 27, 28, 29, 30, 31, 32}.

\section{REALIGNMENT CRITERION: NECESSARY AND SUFFICIENT FOR $X$ STATES}  
In this section, we will show that realignment criterion can be necessary and sufficient criterion for the detection of entangled states, which belongs to a particular large family of two-qubit states, namely $X$-state.\\ 
Let us start with a two-qubit $X$-state described by the density operator $\varrho_{AB}$, which is given by

\begin{equation}
\label{eq:8}
    \varrho_{AB} =
    \begin{pmatrix}
        \rho_{11} & 0 & 0 & \rho_{14} \\
        0 & \rho_{22} & \rho_{23} & 0 \\
        0 & \rho_{23}^* & \rho_{33} & 0 \\
        \rho_{14}^* & 0 & 0 & \rho_{44}
    \end{pmatrix}
    , \rho_{11} + \rho_{22} + \rho_{33} + \rho_{44} = 1
\end{equation}

The non-zero state parameters can be chosen as such that the matrix is positive semi-definite.\\ 
The partial transpose of the density matrix $\varrho_{AB}$ is given by:

\begin{equation}
\label{eq:9}
    \varrho_{AB}^{T_B} =
    \begin{pmatrix}
        \rho_{11} & 0 & 0 & \rho_{23} \\
        0 & \rho_{22} & \rho_{14} & 0 \\
        0 & \rho_{14}^* & \rho_{33} & 0 \\
        \rho_{23}^* & 0 & 0 & \rho_{44}
    \end{pmatrix}
\end{equation}

The eigenvalues of $\varrho_{AB}^{T_B}$ are given as

\begin{equation}
\label{eq:10}
\begin{split}
    \lambda_1 &= \frac{1}{2} \Biggr[(\rho_{11} + \rho_{44}) -  \sqrt{(\rho_{11} - \rho_{44})^2 + 4 |\rho_{23}|^2} \Biggr] , \\
    \lambda_2 &= \frac{1}{2} \Biggr[(\rho_{11} + \rho_{44}) + \sqrt{(\rho_{11} - \rho_{44})^2 + 4 |\rho_{23}|^2} \Biggr], \\
    \lambda_3 &= \frac{1}{2} \Biggr[(\rho_{22} + \rho_{33}) - \sqrt{(\rho_{22} - \rho_{33})^2 + 4 |\rho_{14}|^2}\Biggr], \\
    \lambda_4 &= \frac{1}{2} \Biggr[(\rho_{22} + \rho_{33}) + \sqrt{(\rho_{22} - \rho_{33})^2 + 4 |\rho_{14}|^2}\Biggr].
\end{split}
\end{equation}
A necessary and sufficient condition for entanglement, as established by $PPT$, is that one eigenvalue of $\varrho_{AB}^{T_B}$ is negative, which occurs if either

\begin{equation}
\label{eq:11}
\rho_{22} \rho_{33} < |\rho_{14}|^2 \quad \text{or} \quad \rho_{11} \rho_{44} < |\rho_{23}|^2.
\end{equation}

Since it is known that for $2 \otimes 2$ entangled system, only one eigenvalue of a partially transposed matrix is negative \cite{35a} so only one of the eigenvalues either $\lambda_1$ or $\lambda_3$ given in (\ref{eq:11}) can be negative, but not both. If one condition holds let's say, $\rho_{22} \rho_{33} < |\rho_{14}|^2$ causing $\lambda_3 < 0$, then the other condition must be false, ensuring that $\lambda_1 \geq 0$. Similarly, if $\rho_{11} \rho_{44} < |\rho_{23}|^2$, then $\lambda_1 < 0$ and $\lambda_3 \geq 0$ \cite{35}. Therefore, without loss of generality, we can assume $\lambda_3$ as the negative eigenvalue in the following sections.\\
Now let us perform the realignment operation on the state $\varrho_{AB}$ which transforms it into the following matrix

\begin{equation}
\label{eq:12}
    \mathcal{R}(\varrho_{AB}) =
    \begin{pmatrix}
        \rho_{11} & 0 & 0 & \rho_{22} \\
        0 & \rho_{14} & \rho_{23} & 0 \\
        0 & \rho_{23}^* & \rho_{14}^* & 0 \\
        \rho_{33} & 0 & 0 & \rho_{44}
    \end{pmatrix}
\end{equation}

If $s_1$, $s_2$, $s_3$, $s_4$ denote the singular values of $\mathcal{R}(\varrho_{AB})$ then they can be expressed by the expressions given below:

\begin{equation}
\label{eq:13}
    \scalebox{0.9}{$
    \begin{aligned}
        s_1 &= \frac{1}{\sqrt{2}} \sqrt{ f + \sqrt{(f - 2\rho_{11} \rho_{44} + 2\rho_{22} \rho_{33})
            (f + 2\rho_{11} \rho_{44} - 2\rho_{22} \rho_{33})} } \\
        s_2 &= \frac{1}{\sqrt{2}} \sqrt{ f - \sqrt{(f - 2\rho_{11} \rho_{44} + 2\rho_{22} \rho_{33})
            (f + 2\rho_{11} \rho_{44} - 2\rho_{22} \rho_{33})} } \\
        s_3 &= \frac{1}{\sqrt{2}} \sqrt{ g + \sqrt{(g - 2|\rho_{14}|^2 + 2|\rho_{23}|^2)
            (g + 2|\rho_{14}|^2 - 2|\rho_{23}|^2)} } \\
        s_4 &= \frac{1}{\sqrt{2}} \sqrt{ g - \sqrt{(g - 2|\rho_{14}|^2 + 2|\rho_{23}|^2)
            (g + 2|\rho_{14}|^2 - 2|\rho_{23}|^2)} }
    \end{aligned}
    $}
\end{equation}

where,

\begin{align*}
    f &= \rho_{11}^2 + \rho_{22}^2 + \rho_{33}^2 + \rho_{44}^2, \quad
    g = 2 |\rho_{14}|^2 + 2 |\rho_{23}|^2
\end{align*}

For our convenience, let us denote the the trace norm of $\mathcal{R}(\varrho_{AB})$ as $S$ which will be equal to the sum of singular values. Therefore, we have

\begin{eqnarray}
\label{eq:14}
    || \mathcal{R}(\rho) ||_1 = S \equiv s_1 + s_2 + s_3 + s_4
\end{eqnarray}

Let us now build two small expressions from a large expression of $S$ taking two terms at a time. These small expressions can be called as block and they are given by

\begin{equation}
\label{eq:15}
    P = s_1 + s_2 \text{ and } Q = s_3 + s_4
\end{equation}

Let us first focus on the $P$ block. Using the inequality for sum of squares, we can express the block $P$ as an inequality given by 

\begin{equation}
\label{eq:16}
    s_1 + s_2 \geq \sqrt{s_1^2 + s_2^2}
\end{equation}

Squaring and adding the first two expressions of the singular values given in eq. (\ref{eq:13}), we get  $f = s_1^2 + s_2^2$. Therefore, the inequality (\ref{eq:16}) can be re-expressed as  

\begin{equation}
\label{eq:17}
    s_1 + s_2 \geq \sqrt{f}
\end{equation}

We can also write the inequality in terms of the elements of the density matrix $\varrho_{AB}$ as

\begin{eqnarray}
\label{eq:18}
&& f=\rho_{11}^2 + \rho_{22}^2 + \rho_{33}^2 + \rho_{44}^2 \geq \rho_{11}^2 + \rho_{44}^2  \nonumber\\&&
 \implies \sqrt{f} \geq \sqrt{\rho_{11}^2 + \rho_{44}^2}
\end{eqnarray}

Application of the $AM-QM$ inequality on $\rho_{11}$ and  $\rho_{44}$, we get
\begin{eqnarray}
\label{eq:19}
    \frac{\rho_{11} + \rho_{44}}{2} \leq \sqrt{\frac{\rho_{11}^2 + \rho_{44}^2}{2}}
\end{eqnarray}

Using the inequalities (\ref{eq:17}) and (\ref{eq:18}), the inequality (\ref{eq:19}) reduces to

\begin{equation}
\label{eq:20}
    s_1 + s_2 \geq \frac{\rho_{11} + \rho_{44}}{\sqrt{2}}
\end{equation}

In a similar way, the $Q$ block can be re-expressed as 

\begin{equation}
\label{eq:21}
    s_3 + s_4 \geq \frac{\rho_{22} + \rho_{33}}{\sqrt{2}}
\end{equation}

In terms of the eigenvalues of the partially transposed matrix, the inequalities (\ref{eq:20}) and (\ref{eq:21}) can be written as

\begin{eqnarray}
\label{eq:22}
    s_1 + s_2 \geq \frac{\lambda_{1} + \lambda_{2}}{\sqrt{2}}\geq \sqrt{2}\sqrt{\lambda_1 \cdot \lambda_2}
\end{eqnarray}

\begin{eqnarray}\
\label{eq:23}
    s_3 + s_4 \geq \frac{\lambda_{3} + \lambda_{4}}{\sqrt{2}}
\end{eqnarray}

Since the eigenvalues $\lambda_{1}$ and $\lambda_{2}$ are positive, so we apply $AM-GM$ inequality on $\lambda_{1}$ and $\lambda_{2}$ in the extreme R.H.S of the inequality (\ref{eq:22}).

We consider here through eq. (\ref{eq:23}) that $\lambda_3$ is negative. An alternate situation might be that $\lambda_1$ is considered the negative eigenvalue.

Multiplying the inequalities (\ref{eq:22}) and (\ref{eq:23}), we get 

\begin{equation}
\label{eq:24}
    PQ \geq  \sqrt{\lambda_1 \cdot \lambda_2}\cdot(\lambda_{3} + \lambda_{4})
\end{equation}

Now for two non-negative numbers $P$ and $Q$ with a fixed sum of $S = P + Q$, the product $PQ$ is maximized when $P = Q = \frac{S}{2}$. Therefore, we have

\begin{equation}
\label{eq:25}
    PQ \leq \frac{S^{2}}{4}
    \implies  \frac{S^2}{4} \geq \sqrt{\lambda_1 \cdot \lambda_2}\cdot(\lambda_{3} + \lambda_{4})
\end{equation}

Using (\ref{eq:14}), the inequality (\ref{eq:25}) can be re-written as,

\begin{equation}
\label{eq:26}
    S = || \mathcal{R}(\rho) ||_1  \geq  2 (\lambda_1 \cdot \lambda_2)^\frac{1}{4}\cdot(\lambda_{3} + \lambda_{4})^{\frac{1}{2}}
\end{equation}

\textbf{Theorem-1:}  
A class of two-qubit $X$-states represented by the density operator 
$\varrho_{AB}$ is entangled if and only if
\begin{equation}
\label{eq:27}
	\|\mathcal{R}(\varrho_{AB})\|_1  \geq  2 (\lambda_1 \cdot \lambda_2)^\frac{1}{4} \cdot (\lambda_{3} + \lambda_{4})^{\frac{1}{2}}
\end{equation}

or

\begin{equation}
\label{eq:28}
	\|\mathcal{R}(\varrho_{AB})\|_1  \geq  2 (\lambda_3 \cdot \lambda_4)^\frac{1}{4} \cdot (\lambda_{1} + \lambda_{2})^{\frac{1}{2}}
\end{equation}

accordingly as $\lambda_{3}<0$ or $\lambda_{1}<0$. Here, $\lambda_1$, $\lambda_2$, $\lambda_3$ and $\lambda_4$ denote the eigenvalues of the partially transposed matrix $\varrho_{AB}^{T_{B}}$.\\

\textbf{Corollary-1:}  
Writing the criterion in terms of the elements of the density matrix, the state \(\varrho_{AB}\) is entangled if and only if

\begin{equation}
\label{eq:29}
    \scalebox{0.9}{$
        \begin{aligned}
        \|\mathcal{R}(\varrho_{AB})\|_1 
        &\geq 2 \left[ 
            \left( \frac{1}{2}(\rho_{11} + \rho_{44}) - \frac{1}{2}\Delta_1 \right)
            \left( \frac{1}{2}(\rho_{11} + \rho_{44}) + \frac{1}{2}\Delta_1 \right)
        \right]^{\frac{1}{4}} \\
        &\quad \times \left( \rho_{22} + \rho_{33} \right)^{\frac{1}{2}}.
        \end{aligned}
        $}
\end{equation}

where we define,

\begin{align*}
    \Delta_1 &= \sqrt{(\rho_{11} - \rho_{44})^2 + 4 |\rho_{23}|^2}, \\
\end{align*}

or, 

\begin{equation}
\label{eq:30}
    \scalebox{0.9}{$
        \begin{aligned}
        \|\mathcal{R}(\varrho_{AB})\|_1 
        &\geq 2 \left[ 
            \left( \frac{1}{2}(\rho_{22} + \rho_{33}) - \frac{1}{2}\Delta_2 \right)
            \left( \frac{1}{2}(\rho_{22} + \rho_{33}) + \frac{1}{2}\Delta_2 \right)
        \right]^{\frac{1}{4}} \\
        &\quad \times \left( \rho_{11} + \rho_{44} \right)^{\frac{1}{2}}.
        \end{aligned}
        $}
\end{equation}

where we define,

\begin{align*}
    \Delta_2&= \sqrt{(\rho_{22} - \rho_{33})^2 + 4 |\rho_{14}|^2}, \\
\end{align*}

\section{ENTANGLEMENT DETECTION IN A GENERALISED TWO-QUBIT $X$-STATE}

In this section, we analyse an example of a generalised two-qubit $X$-state that remains undetected by the realignment criterion but can be detected using Theorem-1. We define constraints on its parameters such that all the density matrices of this form within those constraints satisfies Theorem-1 and thus the density matrices are entangled.\\
To elucidate our discussion, we consider a particular form of the $X$-state, which is described by the density operator $\varrho_1$

\begin{equation}
\label{eq:31}
    \varrho_1 = 
    \begin{pmatrix}
        0.35 & 0 & 0 & x \\
        0 & 0.25 & y & 0 \\
        0 & y & 0.25 & 0 \\
        x & 0 & 0 & 0.15
    \end{pmatrix}.
\end{equation}

First, we impose the requirement that $\varrho_1$ must be positive semidefinite, ensuring non-negativity of its eigenvalues. This imposes constraints on the state parameter $x$ and $y$. Using the formulas from \cite{35}, the eigenvalues of $\varrho_1$ are:

\begin{equation}
\label{eq:32}
    \begin{aligned}
        \lambda_1 &= 0.25 + \sqrt{x^2 + 0.01}, \quad
        \lambda_2 = 0.25 - \sqrt{x^2 + 0.01}, \\
        \lambda_3 &= 0.25 + y, \quad
        \lambda_4 = 0.25 - y.
    \end{aligned}
\end{equation}

Enforcing $\lambda_i \geq 0$ for all $i=1,2,3,4,$ leads to the following constraints

\begin{equation}
\label{eq:33}
    0 < x < 0.2291, \quad 0 < y < 0.25
\end{equation}

Now, our task is to inspect the different entanglement criterion such as partial transposition criterion, realignment criterion and modified realignment criterion for the detection of entanglement in $\varrho_1$. This study helps us to investigate the efficiency of the modified realignment criterion over partial transposition and realignment criterion.\\

\textbf{ (i) Partial Transposition Criterion:} In this criterion, we consider the partial transpose of the density matrix $\varrho_1$, which is given by

\begin{equation}
\label{eq:34}
    \varrho_1^{T_B} = 
    \begin{pmatrix}
        0.35 & 0 & 0 & y \\
        0 & 0.25 & x & 0 \\
        0 & x & 0.25 & 0 \\  
        y & 0 & 0 & 0.15
    \end{pmatrix}.
\end{equation}

The eigenvalues of $\varrho_1^{T_B}$ are:

\begin{equation}
\label{eq:35}
    \begin{aligned}
        \lambda_1' &= 0.25 - \sqrt{y^2 + 0.01}, \quad
        \lambda_2' = 0.25 + \sqrt{y^2 + 0.01}, \\
        \lambda_3' &= 0.25 - x, \quad
        \lambda_4' = 0.25 + x.
    \end{aligned}
\end{equation}

For $\varrho_1$ to be detected as entangled via the partial transposition criterion, one of the eigenvalues of $\varrho_1^{T_B}$ must be negative. This can be studied in two possible scenarios in which either $\lambda_1' < 0$ or $\lambda_3' < 0$:

\begin{itemize}
    \item \textbf{Scenario-A:} If $\lambda_1' < 0$ while all other eigenvalues remain positive, we obtain the bound as

    \begin{equation}
    \label{eq:36}
        y > 0.2291.
    \end{equation}

    Additionally, since $\lambda_3' > 0$ implies $x < 0.25$, which is already satisfied by \eqref{eq:33}, no further constraints on $x$ arise. Therefore, in this scenario, the state $\varrho_1$ is entangled if and only if
    
    \begin{equation}
    \label{eq:37}
     0 < x < 0.2291, \quad 0.2291 < y < 0.25.
    \end{equation}

    \item \textbf{Scenario-B:} If $\lambda_3' < 0$ while all other eigenvalues remain positive, then $x > 0.25$, which contradicts bound in eq. \eqref{eq:33}. Thus, this case does not hold good.
\end{itemize}

\textbf{(ii) Realignment Criterion:} Let us now analyse the realignment criterion. Applying the realignment operation to $\varrho_1$ gives the following matrix:

\begin{equation}
\label{eq:38}
    \mathcal{R}(\varrho_{1}) =
    \begin{pmatrix}
        0.35 & 0 & 0 & 0.25 \\
        0 & x & y & 0 \\
        0 & y & x & 0 \\
        0.25 & 0 & 0 & 0.15
    \end{pmatrix}.
\end{equation}
\\
The realignment criterion detects entanglement in $\varrho_{1}$ if the following condition holds:

\begin{equation}
\label{eq:39}
    \|\mathcal{R}(\varrho_1)\|_1 >1.
\end{equation}

Explicitly, this translates to:

\begin{equation}
\label{eq:40}
    \sqrt{x^2 - 2xy + y^2} + \sqrt{x^2 + 2xy + y^2} + 0.5385 > 1
\end{equation}

The inequality (\ref{eq:40}) can be solved by considering two cases, which are given below:

\begin{itemize}
    \item \textbf{Case-1:} If $x > y$ then solving the inequality (\ref{eq:40}), we get
    \begin{equation}
    \label{eq:41}
        x > 0.2307
    \end{equation}
    \item \textbf{Case-2:} If $x < y$ then the inequality (\ref{eq:40}) gives
    \begin{equation}
    \label{eq:42}
        y > 0.2307
    \end{equation}
\end{itemize}

Thus, Case-I can be eliminated as the obtained bound of $x$ violates the inequality in Eq. (\ref{eq:33}). 

Therefore, entanglement can be detected via the Realignment Criterion when:

\begin{equation}
\label{eq:43}
    0 < x < 0.2291, \quad 0.2307 < y < 0.25
\end{equation}

Finally, we analyse the modified realignment criterion stated in Theorem-1. Since $\lambda_1'$ is negative in this case, we plug in the trace norm of $\varrho_1$ and the eigenvalues of $\varrho_1^{T_B}$ into Theorem-1, specifically eq. (\ref{eq:28}). We get the inequality given below:

\begin{equation}
\label{eq:44}
    |x-y| + |x+y| + 0.5385 \ge 2\big[0.0625 - x^2]^{\frac{1}{4}}\big[0.5]^{\frac{1}{2}}
\end{equation}

If we consider $y > x$ then the above inequality reduces to

\begin{equation}
\label{eq:45}
    2y + 0.5385 \ge \sqrt{2}\big[0.0625 - x^2]^{\frac{1}{4}}
\end{equation}

Solving the inequality (\ref{eq:45}) for $y$ and express the solution in terms of $x$, we get

\begin{equation}
\label{eq:46}
    y \geq \frac{\sqrt{2}\cdot(0.0625 - x^2)^{\frac{1}{4}} - 0.5385}{2}
\end{equation}

Let us denote the right-hand side of eq.~(\ref{eq:46}) by \( f(x) \). It is evident that \( f(x) \) is a monotonically decreasing function in $0<x<0.2291$. Consequently, the maximum value of \( f(x) \) occurs as \( x \to 0 \). Since eq.~(\ref{eq:46}) provides an upper bound on the expression involving \( y \), we evaluate \( f(x) \) in the limit as \( x \to 0 \) to obtain a conservative estimate:

\begin{align}
\label{eq:47}
    \text{At } x = 0,\quad f(0) &= 0.08430 \nonumber \\
    \implies\quad y &\geq 0.08430
\end{align}

Since the lower bound of $y$ given in eq. (\ref{eq:37}) is significantly greater than the upper bound in eq. (\ref{eq:46}), the inequality holds for all $x$ and $y$ defined using PPT.

Therefore the state $\varrho_1$ will be entangled if and only if,

\begin{equation}
 0 < x < 0.2291 \text{ and } 0.2291 < y < 0.25    
\end{equation}

It is observed here that the bound obtained via the Modified Realignment Criterion exceeds that obtained through the Realignment Criterion in $y$. Consequently, this indicates that our criterion is capable of detecting a greater number of states.

\section{CONCLUSION}
This study has provided a comprehensive review of the realignment criterion, a necessary but not sufficient condition for detecting entanglement in both low- and high-dimensional quantum systems. To address its limitations, we have refined this criterion for a specific class of two-qubit states, known as $X$-states, by introducing a state-dependent entanglement detection framework based on the realignment operation. Unlike the general realignment criterion, which is state-independent, our refined approach leverages state-specific properties to enhance its effectiveness. It is important to note that, while we have derived a necessary and sufficient condition using the realignment operation, this derivation comes at the cost of having to satisfy the Partial Transposition Criterion in order to determine which eigenvalue is taken to be negative. The development of a criterion that is both necessary and sufficient for entanglement detection in general $d \otimes d$ systems, however, still remains an open challenge.

\section{DATA AVAILABILITY STATEMENT}

Data sharing is not applicable in this article as no datasets were generated or analysed during the current study. 

\bibliography{refs}

\end{document}